**A heterogeneous lithosphere on Venus**

I. Romeo[a]*, A. Jiménez-Díaz[b], M. Mendiburu-Eliçabe[c], I. Egea-González[d], R.M. Hahn[e], P.K. Byrne[e], J. F. Kirby[f], J. Álvarez-Lozano[a] & J. Ruiz[a]

a) Departamento de Geodinámica, Estratigrafía y Paleontología. Universidad Complutense de Madrid. 28040 Madrid, Spain.

b) Departamento de Biología y Geología, Física y Química Inorgánica, ESCET, Universidad Rey Juan Carlos, Móstoles, Madrid, 28933, Spain.

c) Departamento de Estadística e Investigación Operativa. Universidad Complutense de Madrid. 28040 Madrid, Spain.

d) Departamento de Física Aplicada, Escuela Superior de Ingeniería, Universidad de Cádiz, 11519, Puerto Real, Cádiz, Spain.

e) Department of Earth, Enviromental and Planetary Sciences, Washington University in St. Louis, St. Louis, MO 63130, USA.

f) Geodesy & Earth Observation, DTU Space, Technical University of Denmark, 2800 Kongens Lyngby, Denmark.

* Corresponding author: iromeobr@ucm.es  +34 91 394 4821

**Venus is 95% the size of Earth and probably has a similar composition and internal energy, but the geodynamical mechanism governing its internal thermal evolution remains controversial. Much of the planet surface (77%) is covered by volcanic plains, characterized by basaltic flood lavas, some of which emanate from volcanoes whereas others have no identifiable sources. Here we show that the volcanic plains contain three provinces with statistically different properties. Apart from the plains around the Beta-Atla-Themis region characterized by a thick crust with a wide range of effective elastic lithospheric thickness ($T_e$),**

**the remaining volcanic plains (61% of the surface) display a dichotomous nature, with the northern plains showing a thinner lithosphere and twice the number of volcanoes per unit area compared to the southern plains. These global-scale differences imply a complex geodynamical regime, with large lateral variations in mechanical properties and/or geodynamical processes, which must be addressed by competing models of mantle convection and planetary evolution.**

The current geodynamical regime on Venus, a consequence of the planet's evolution principally as a function of heat loss, is the subject of intense debate (1-6). Geodynamic models must explain a distributed style of tectonic deformation (no as-yet recognized plate boundaries) with a well-developed branched rift system between active mantle upwellings in the Beta, Atla, Themis (BAT) Region (7), together with limited compressional deformation belts (8) and some strike-slip deformation zones (9-11). The mechanical properties of the lithosphere of the volcanic plains occupying much of the planet surface, including its lateral variations, have been necessarily shaped by the geodynamic history of Venus.

The planet show ~942 impact craters (12) with a spatial distribution indistinguishable from random (13, 14). The low proportion of impact craters clearly modified by volcanism (6-13%) suggests a nearly global resurfacing event where most previous craters were removed, followed by a period of limited geological activity (3, 15-17). Nevertheless, other studies provide higher estimates of volcanically embayed impact craters (18) which supports the idea that spatial randomness can be maintained through equilibrium resurfacing by small patches of volcanic plains (19). However, the frequency-size distribution of the mapped volcanic units does not show the abundance of small patches required by the equilibrium resurfacing hypothesis (3, 20). A global resurfacing event can be caused by different geodynamical processes, including nearly global subduction of the lithosphere a behavior shown by models of mantle convection with episodic lithospheric overturns (5,6,21), global volcanic resurfacing driven by a transition from mobile to stagnant lid (2), or global volcanism driven by the rupture of the basalt barrier in the mantle transition zone causing global episodic mantle overturns in stagnant-lid (22). Limited lateral movements of lithospheric blocks have been reported in the volcanic plains (23), indicating that the lithosphere is not completely stagnant. A "plutonic squishy lid" regime allows the observed lateral lithospheric movements (24). A "deformable episodic lid" regime has been achieved in convection models incorporating a weak plagioclase-rheology crust, resulting in lithospheric overturns separated by periods of mobile lid behavior (6). 3D convection models show complex evolutions with coexisting portions of the lithosphere exhibiting different concurrent geodynamic regimes (25) (e.g. part of the planet behaves as if it were in stagnant lid mode with other parts showing episodic lid overturns). Tessera terrain, locally the oldest unit in terms of relative age and characterized by high-angle, cross-cutting tectonic structure sets, has been proposed to represent differentiated crust, sufficiently low density to survive global lithospheric subduction events (26). A felsic composition for tessera terrain has been favored by infrared emissivity observations (27-29) and structural studies (30, 31). Convection models indicate that some portions of the lithosphere can withstand lithospheric mantle overturns (5, 21).

A key characteristic of the Venus lithosphere that mantle-convection models must address is its thickness and lateral spatial variations, which is analyzed in this study. To accomplish this, we used a version of the effective elastic lithospheric thickness model in (32), calculated by inverting the spectral relations between gravity and topography using the wavelet method. Additionally, we analyzed the crustal thickness model in (32). We spatially categorized the geophysical data including the effective elastic lithospheric thickness, $T_e$, (Fig.1a), the crustal thickness, $T_c$, (Fig.1b) and the subsurface load fraction, $F$, (Fig.1c) according to interpreted global geological provinces (Fig.1d). $T_e$ is a proxy for the strength of the lithosphere, integrating contributions from brittle and ductile layers and from elastic cores of the lithosphere (33). $F$ indicates the ratio between internal and surface loads on the lithosphere, ranging from 0, only surface loads, to 1, only internal loads. Here, we analyze the data from the volcanic plains, because they represent the largest geological unit on the planet where we can observe the average characteristics of the Venusian lithosphere excluding local geodynamical processes in active areas (rifts and volcanic rises) or ancient terrains (tessera). We found that the lithosphere of the volcanic plains is not spatially homogeneous in terms of $T_e$, and $F$, so we delineate three large units (Fig.1d) that are found to be statistically significantly different. Finally, we analyzed the differences in volcano spatial density among these units confirming strong lateral variations in the Venusian lithosphere.

**RESULTS**

**Crustal and lithospheric characteristics of the volcanic plains**

The data of the analyzed parameters ($T_e$, $T_c$, and $F$) for the volcanic plains shows statistically significant geographical variations. The BAT region shows a thicker crust (2 km over the average) with higher $T_e$ values (15 km over the average) and $F$ values indicating external loads (Fig. 2). The rest of the volcanic plains, which represents 61% of the planet surface, are not spatially homogeneous, exhibiting a dichotomous distribution of $T_e$ values. The northern plains present effective elastic thicknesses that are ~8 km lower than the plains in the planet's southern hemisphere (Fig. 2d). The boundary between both domains is not sharp, featuring a gradient. The BAT region together with the Dali–Diana–Artemis–Juno Chasmata along southern Aphrodite Terra are located between both domains. Therefore, there is only a location where the northern and southern plains contact. This limit lies between Aino and Tahmina planitiae, and continues between Lada Terra and Alpha Regio towards Parga Chasma, providing a statistically significant difference in $T_e$ between both data sets (Fig.2). Moreover, the northern plains exhibit a predominance of internal loads whereas the southern plains present an equilibrium between internal and external loads; this difference in the $F$ parameter between these two regions is also statistically significant (Fig.2). The differences in $T_e$ values between the hemispherically regional plains units can also be recognized in other studies of Venus' effective elastic lithospheric thickness (34, 35), although it passed unnoticed so far.

**Statistics**

The data population of $T_e$, $T_c$, and $F$ does not follow a normal distribution; therefore, we performed a Kruskal–Wallis test between the three volcanic plains provinces we consider here, the BAT, northern, and southern plains (Fig.2 d,e,f) in order to establish whether they belong to the same statistical population or exhibit significant differences. The extremely low *p*-values ($1.95 \times 10^{-307}$ for $T_e$, $5.92 \times 10^{-268}$ for $T_c$ and $1.49 \times 10^{-156}$ for $F$) provide robust evidence indicating that the three provinces show strong differences among them. After the positive result obtained from the Kruskal–Wallis test, we carried out a post-hoc Dunn test for each pair of units and each parameter. The *p*-values of this post hoc analysis are shown in Table 1. The differences in $T_e$ between of the northern and southern plains yield a very low *p*-value, indicating a clear difference in lithospheric thickness. The comparison between the three major plains units shows a similarly strong difference for $T_e$ and $T_c$. Nevertheless, the difference between the southern and BAT plains is stronger for the $F$ value, which is characterized by external loads on the BAT plains because of the presence of numerous volcanoes.

Since the special nature of the geological features of the BAT area is well established (17, 36, 37), we focused on the analysis of the data from the volcanic plains outside the BAT area. We built up a fuzzy model to automatically uncover statistical clusters based on geophysical values in the volcanic plains outside of the BAT area, which result agrees with our initial delineation of the northern and southern plains. The model automatically separates data as a function of $T_e$ and $T_c$. The model identified the existence of two clusters of data, the spatial distribution of which is shown in Fig. 3a. The model considered $T_e$ values to be the most efficient way to separate both clusters (Fig. 3c), which was not initially prescribed in the model setup. Cluster 1 dominates in the northern plains and Cluster 2 dominates in the southern plains (Fig. 3b). Extremely low *p*-values obtained from a chi-square test confirm the different nature of the lithosphere of the northern and southern plains based on $T_e$.

**Spatial distribution of volcanoes and coronae**

The detailed volcano catalog compiled in (38) including ~85,000 volcanic centers, allows an accurate evaluation of the preserved volcanic activity in each of the three proposed volcanic plains provinces. The concentration of large volcanoes in the BAT anomaly area was pointed out by (36). This BAT concentration is also evident for small volcanoes. The volcanic plains in the BAT anomaly area (excluding the large volcanic rises of Beta, Atla, Themis, and the rift systems between them) show the highest spatial concentration of volcanoes <5 km (Table 2). When examining the volcano spatial density in the northern and southern plains (defined in this study by differences in effective elastic lithospheric thickness) we found that the northern plains (thinner lithosphere) have twice the volcano spatial density of the southern plains (Table 2, Fig.4). The spatial density of volcanoes <5 km in the northern plains is 207% of those found in the south (more than double of volcanoes per unit area). This proportion for volcanoes from 5 to 100 km in diameter is 126%. For the largest volcanoes ( >100 km in diameter) this value is 243%, indicative of a strong concentration of large volcanoes in the northern plains. Nevertheless, the spatial density of the volcanotectonic corona landforms is very similar between the northern and southern plains (Table 2).

**DISCUSSION**

**Implications of a heterogeneous lithosphere in the geodynamical regime**

Despite recent convection models (5, 6, 21) do not support a present stagnant lid regime for Venus, geological studies advocate for limited volcanic and tectonic activity at present (2, 4, 39). There is discussion about whether the amount of heat that can be transferred by conduction and magmatic advection across an immobile lithosphere is sufficient to balance the heat generated by radiogenic production in Venus' mantle. If the heat lost by conduction and magmatic advention is insufficient, this leads to mantle heating and a possible destabilization of the lithosphere, promoting the cessation of the stagnant-lid regime. The answer is complex, as it depends on unknown facts like the efficiency of Venus' history of planetary cooling and the amount of radioactive elements concentrated in the crust (which are also conditioned by the geodynamical regime and its evolution with time).

The effect of active mantle upwellings connected by rift systems in the BAT region affects the characteristics of the lithosphere of the volcanic plains in this area, providing a thick crust and *F* values indicating external loads (Fig. 2). All these observations are consistent with a crustal thickening by volcanic materials generating loads on the lithosphere, which agree with the concentration of volcanoes in this area (36) driven by the presence of active mantle upwellings. The difference of effective elastic thicknesses between the northern and southern plains can be related to variations in the thermal state of these regions (40). If we use the methodology describing this relation (40), a dry mantle rheology, a surface temperature of 740 K, and the mean effective elastic thicknesses of the lithosphere at the northern and southern plains (36.4 and 44.0 km respectively), then the corresponding average thermal gradients would be, respectively, 12 and 10 K km$^{-1}$. Thus, the heat flow when the gravity/topography relation of these regions was established was about 20% higher in the northern plains compared to the southern plains. Possible explanations for the hemispheric lithospheric heterogeneity we find include: 1) the northern and southern plains have different ages and thus different lithospheric thicknesses controlled by lithospheric cooling through conduction; 2) they are subject to, and operate within, different but contemporary geodynamical regimes; 3) they behave within the same geodynamical regime but with different levels of activity or 4) a complex combination of these, i.e. they have experienced different geodynamical histories. These possibilities are consistent with recent numerical convective models, the survival of portions of the lithosphere following global subduction events accounts for possible differences in age (5, 21). Similarly, contemporary areas with different geodynamical regimes were replicated in 3D convective models (25).

**The role played by tessera terrain**

The spatial relation with tessera terrain is different between the northern and southern plains. The northern plains surround all the main tessera occurrences (Ovda, Thetis, Tellus, Fortuna and Alpha) and contain the majority of small tessera inliers. However, the southern plains do not include any major tessera occurrence, and tessera inliers are relatively scarce. The effective elastic lithospheric thickness of tessera terrains is typically less than their crustal thicknesses (32), probably registering the mechanical state of the lithosphere at the time when the last tectonic deformation occurred. Since the northern plains surround the main tessera occurrences and host numerous tessera inliers, which presumably indicates that these plains were emplaced on spatially extensive tessera units, it can be deduced that the northern plains' lower $T_e$ values are affected by and reflect the signature of underlying tessera terrain units.

**Formation age as a possible factor controlling lithospheric thickness**

Absolute age determinations by crater counting on Venus are biased by the scarce population of impact craters (41, 42). For crater counting, we used the third release of the crater database of (43). The northern plains have 438 craters in $201.34 \times 10^6$ km², which corresponds to a crater spatial density of $2.18 \times 10^{-6}$ craters/km². The southern plains have 165 craters in $79.47 \times 10^6$ km², which corresponds to a crater spatial density of $2.08 \times 10^{-6}$ craters/km². We have performed a statistical Monte Carlo simulation where craters are accumulated randomly in space and time on both units with distinct age difference between the northern and southern plains (Fig. 5). 10,000 simulations were performed for each increment of 0.001 in a relative age difference line from -1 to 1, where 0 represents the same age for both units, -1 represents that the southern plains were present during all the model time span but the northern plains have just formed, and 1 represents the opposite scenario. This stochastic model considering a 95% confidence interval provides an age ratio range of -0.13–0.20, which means that both units could have the same age but in the extreme cases the northern plains could be 13% younger than the southern plains, or the southern plains could be 20% younger than the northern plains. Therefore, although the northern plains have a higher crater spatial density, this difference in crater statistics could be due not to an older age but to the stochastic variability of the cratering process. The average crater retention age of the Venusian surface is not well understood, estimated to range from 1000 to 300 Ma (41). Taking 750 Ma for the southern plains, the crater statistics allow an age of even 97.5 Ma younger for the northern plains (-0.13 age ratio). Thus, we cannot rule out that the observed difference of lithospheric thickness is due to a lithosphere formation age difference, with a lithosphere that thickened by conduction since its formation (in the case where surface age corresponds approximately to the time of lithosphere formation). At the opposite limit of the 95% confidence interval, taking 750 Ma for the northern plains, the crater statistics allow an age of 150 Ma younger for the southern plains (0.2 age ratio).

**Lateral variations of the geodynamical regime**

The other fundamental source of lithospheric heterogeneity apart from age differences is the coexistence of distinct geodynamical regimes and/or geodynamical histories.

On the one hand, the thinner lithosphere of the northern plains can be maintained by a plutonic squishy-lid regime (24) or deformable episodic lid (6), where most of the heat is transferred by intrusive materials in the crust, together with limited lithospheric delamination. This regime is compatible with the observations of limited lateral lithospheric movements found in numerous places across numerous plains sites on Venus (23). On the other hand, the thicker lithosphere of the southern plains could currently mainly behave as a stagnant-lid regime, the stability of which is maintained by a thicker and more rigid lithosphere. Supporting this interpretation, most of the reported occurrences of lateral lithospheric movements are located within the northern plains or along the boundary of the southern plains.

The difference in volcano spatial density between the northern and southern domains also indicates a strong difference in the importance of magmatic advection as a heat transfer mechanism through the lithosphere. The thin lithosphere in the north correlates with a high spatial concentration of volcanoes of all the sizes (Table 2, Fig.4.), indicating that both conduction and magmatic advection are easier through a thin lithosphere. This difference in lithospheric thickness appears to control the formation of volcanic edifices, but seemingly has no influence on corona spatial density (Table 2). If corona are the volcanotectonic manifestation on the surface of mantle upwellings impinging under the lithosphere (44), then corona occurrence is mainly controlled by mantle dynamics but not by lithospheric thickness. Confirming the role of the mantle in corona generation, corona activity is not related with Te, finding both active/inactive coronae (45) in the northern thin and southern thick lithosphere.

**Conclusion**

The Venus lithosphere is heterogeneous at a global scale. There is a statistically significant difference in lithospheric thickness between the northern and southern volcanic plains beyond the BAT region. The relatively thinner lithosphere in the northern hemisphere hosts approximately twice the volcano spatial density than that in the south. This is a key physiographic characteristic of the Venusian lithosphere that future models of geodynamical evolution of the planet must address.

**METHODS**

**Crustal thickness ($T_c$), effective elastic thickness ($T_e$) and subsurface load fraction ($F$) data**

We use the crustal thickness ($T_c$) model from (32). The data of the effective elastic thickness ($T_e$) and subsurface load fraction ($F$) were obtained from an updated version of the model of Jiménez-Díaz et al. (32), averaged onto 2º × 2º grids. The original inversion for $T_e$ was performed on observed coherences with wavelengths >211 km, whereas the updated inversion includes the complete range of available wavelengths, which data are available at https://doi.org/10.5281/zenodo.13138657. For details of the inversion procedure see (32). The data of $T_c$, $T_e$, and $F$ models were sampled with a homogeneous spatial density of points (maintaining the number of data points per unit area constant across all the latitude range) with a resolution of 211.22 km, equal to the resolution of the sampled models. Finally, these data were classified using a map of geological provinces. The data set is available at https://doi.org/10.5281/zenodo.15719877.

**Map of geological provinces**

The global map of geological provinces was produced for this study using NASA Magellan data altimetry and full resolution (75-100 m /pixel) S-band SAR imagery in QGIS. Large tessera areas (Ovda, Tethis, Haaste Bad, Alpha, Fortuna, Tellus, Phoebe, Thetus, Doyla), mountain ranges (Maxwell, Akna, Freyja), Lakshmi Planum, volcanic rises (Beta, Atla, Themis, Idmr, Bell, Eistla, Dione, Lada), rift belts (Devana, Ralk-Umgu, Dali, Parga, Ganis, Hecate, Artemis) and the regional volcanic plains were mapped. Geological mapping was mainly performed in QGIS using a wiew scale of 1:5000000, but a more detailed scale was used when needed. The limits of tessera terrains can be accurately delineated in most cases since the overlapped volcanic plains embay their margins. Only the main outcrops of tessera terrain were mapped. Numerous tessera inliers present inside the volcanic plains are included in the volcanic plains units. The limits of rift belts were traced at the last observed extensional lineament interpreted to belong to that belt. The limits of volcanic rises were delineated based on where the main topographic signature is lost, where radial volcanic flows end, or where the tectonic structures associated with the volcanic rise (radial or concentric) disappear. The discrimination between the volcanic plains units was performed attending to the variation of $T_e$ and $F$. The global map of geological provinces is available at https://doi.org/10.5281/zenodo.15719877.

**Statistical methods**

We examined whether there were statistically significant differences between the variables $T_e$, $T_c$, and $F$ across three geological provinces of the volcanic plains: the northern, southern, and BAT plains. Normality and homogeneity of the data were assessed using the Shapiro–Wilk and Levene's tests. Given the non-normality and unequal variances observed, we applied the non-parametric Kruskal–Wallis test to compare the groups. Following the significant results from the Kruskal–Wallis test, a post-hoc analysis using Dunn's test with Bonferroni correction was conducted to assess which specific group pairs exhibited significant differences.

Given the novelty of our differentiation between the northern and southern plains, we constructed an automatic statistical model to identify clusters of data in the volcanic plains outside the BAT area in order to test if our separation between these hemisphere-scale plains was statistically justified. Fuzzy clustering was performed in our analysis using the fanny algorithm from the R package to divide the data into clusters based on $T_e$ and $T_c$. The fuzzy clustering method was used to assign the membership of data points across clusters. The number of cluster was not prescribed by the method. The quality of the clustering was evaluated using the Dunn partition coefficient. After clustering, data points were assigned to their most dominant cluster. The results were visualized in scatter plots, with points color-coded by cluster membership, to assess the relationships between $T_e$ and $T_c$. The fuzzy clustering method was successfully applied to geophysical data correlation (e.g. in 46,47) and has also proven useful for automated geological mapping (48-50).

**Spatial density measurements of volcanoes, coronae and impact craters**

We used the volcano catalog compiled in (38) including ~85,000 volcanic centers, the corona catalog of (51) merged with the corona database of (45) and the third release of the impact crater database of (43). Object counting for each data set on each unit was performed using QGIS. In order to calculate the spatial densities shown in Table 2, we used the area of each unit subtracting the small gaps of Magellan radar imagery where volcanoes, coronae, and impact craters cannot be identified.

<div style="text-align:center">**REFERENCES**</div>

# Tables

**Table 1. Adjusted P-values of the comparison between pairs of units with Dunn post hoc test**

| Unit comparison | $T_e$ | $T_c$ | $F$ |
|---|---|---|---|
| North Plains - South Plains | $1.70 \cdot 10^{-111}$ | $9.98 \cdot 10^{-20}$ | $5.15 \cdot 10^{-66}$ |
| North Plains - BAT Plains | $2.83 \cdot 10^{-269}$ | $4.27 \cdot 10^{-269}$ | $8.46 \cdot 10^{-51}$ |
| South Plains - BAT Plains | $5.05 \cdot 10^{-29}$ | $8.60 \cdot 10^{-108}$ | $9.41 \cdot 10^{-157}$ |

**Table 2. Counting and spatial density of impact craters, coronae and volcanoes**

| | | Impact Craters | | Coronae | | Volcano < 5km | | Volcano 5km-100 | | Volcano > 100 km | | Volcanic Fields | |
|---|---|---|---|---|---|---|---|---|---|---|---|---|---|
| Unit | Area (km²) | N | N/km² | N | N/km² | N | N/km² | N | N/km² | N | N/km² | N | N/km² |
| North Plains | 192511630 | 438 | $2.28 \cdot 10^{-6}$ | 146 | $7.25 \cdot 10^{-7}$ | 47301 | $2.46 \cdot 10^{-4}$ | 325 | $1.69 \cdot 10^{-6}$ | 38 | $1.97 \cdot 10^{-7}$ | 284 | $1.48 \cdot 10^{-6}$ |
| South Plains | 73966292 | 165 | $2.23 \cdot 10^{-6}$ | 58 | $7.30 \cdot 10^{-7}$ | 8821 | $1.19 \cdot 10^{-4}$ | 99 | $1.34 \cdot 10^{-6}$ | 6 | $0.81 \cdot 10^{-7}$ | 68 | $0.92 \cdot 10^{-6}$ |
| BAT Plains | 70219895 | 139 | $1.98 \cdot 10^{-6}$ | 55 | $7.69 \cdot 10^{-7}$ | 19406 | $2.76 \cdot 10^{-4}$ | 165 | $2.35 \cdot 10^{-6}$ | 35 | $4.98 \cdot 10^{-7}$ | 149 | $2.12 \cdot 10^{-6}$ |

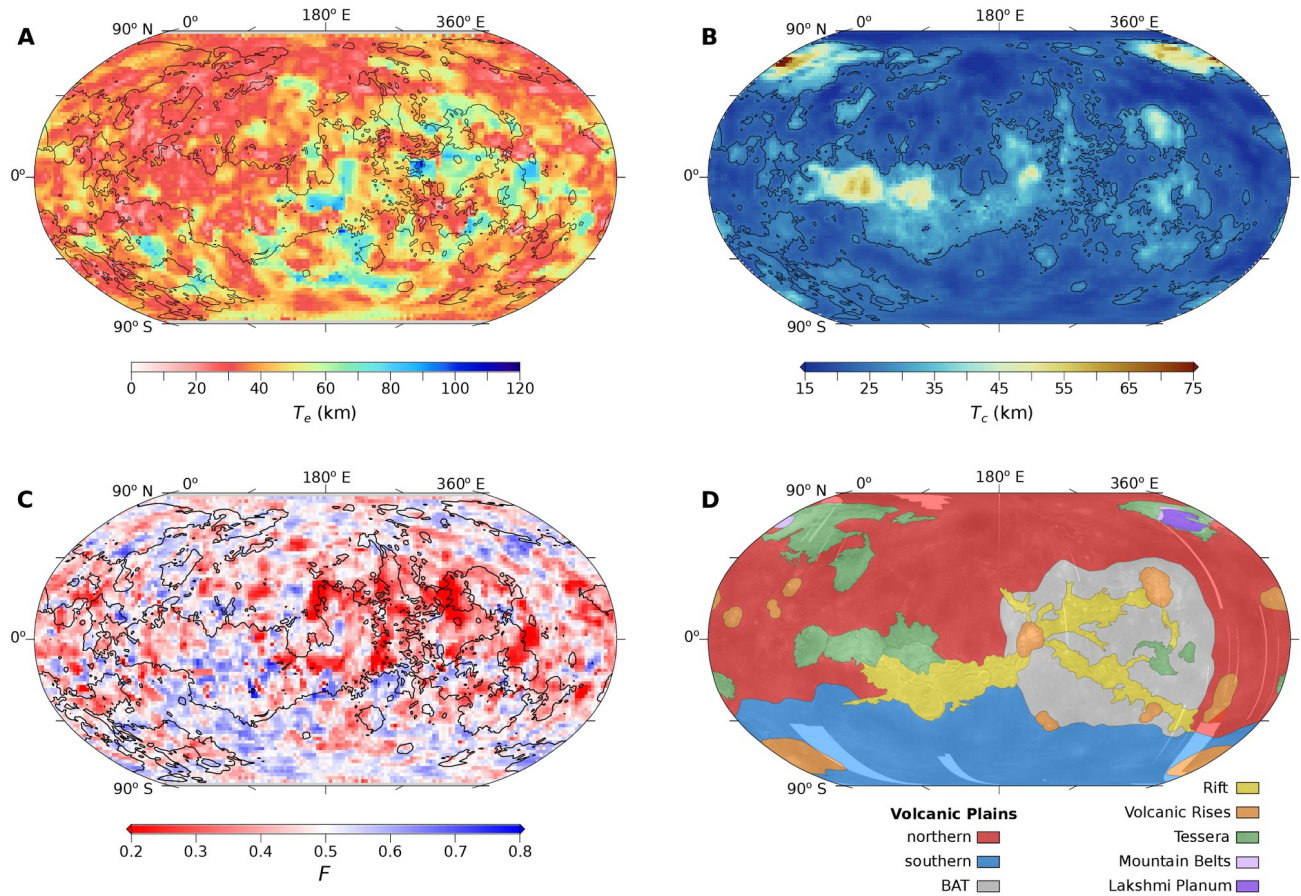

**Fig. 1. Geophysical data sets and geological map used.** (**A**) Global map of the effective elastic lithospheric thickness ($T_e$) with a resolution of 2° × 2°; the total $T_e$ range is between 4.8 and 105 km. (**B**) Global map of the crustal thickness ($T_c$) with an average value of 25 km. (**C**) Global map of the ratio between internal and surface loads (*F*) on the lithosphere from 0 (surface loads) to 1 (internal loads). (**D**) Global map of the geological provinces used to classify the data set. The three volcanic plains units were separated using $T_e$ and *F* variation. The four maps use the Robinson projection.

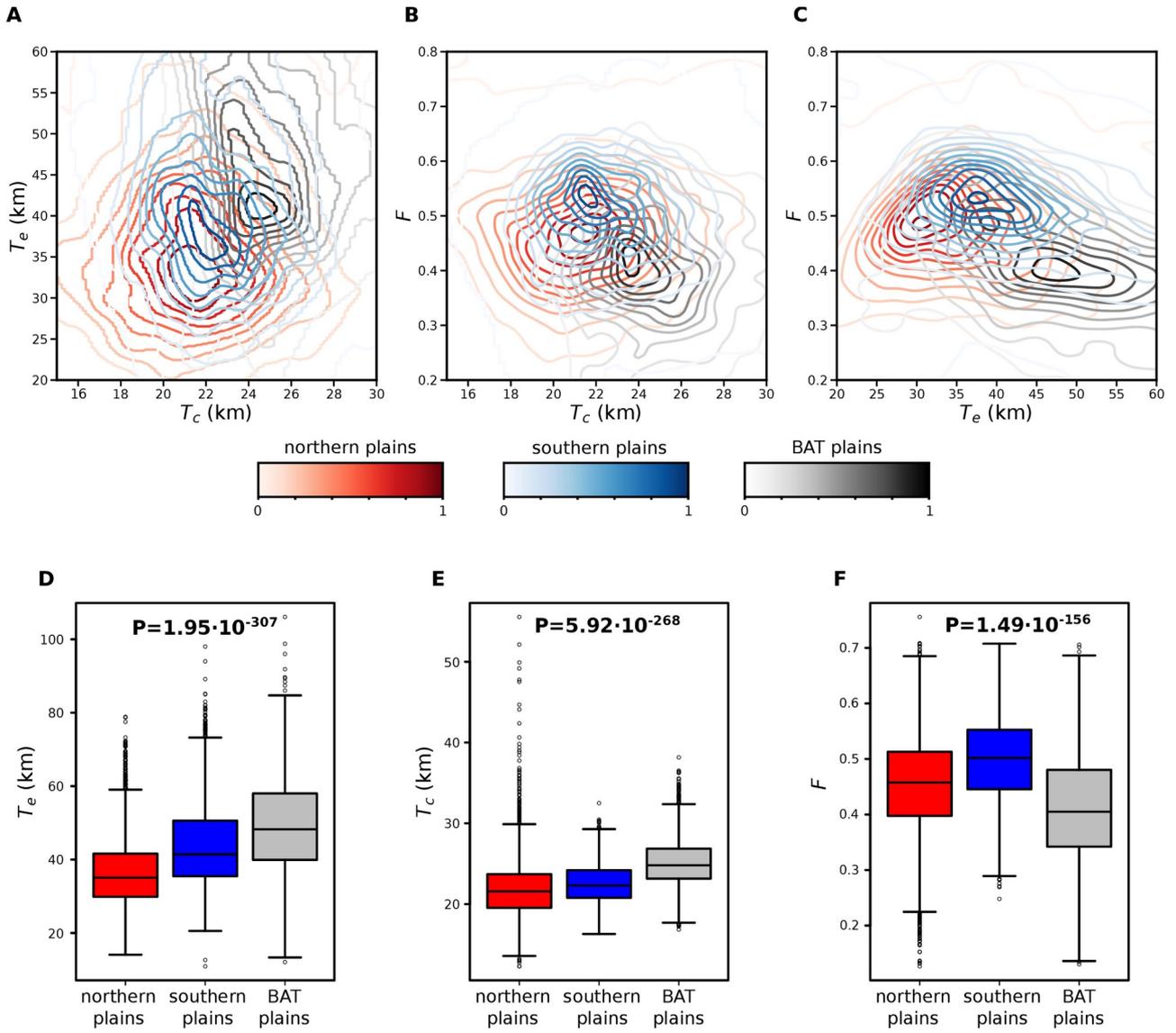

**Fig. 2. Volcanic plains lithospheric characteristics.** Effective elastic lithospheric thickness ($T_e$), crustal thickness ($T_c$), and subsurface load fraction ($F$) of the volcanic plains on Venus, classified into three provinces: northern plains, southern plains, and BAT. The parameter maps comparing different pairs of parameters are shown in (**A**), (**B**), and (**C**). Color bars indicate normalized data density (from 0 indicating minimum data density to 1 indicating maximum data density). Box plots of the data distributions comparing different volcanic plains provinces are shown in (**D**), (**E**) and (**F**) for $T_e$, $T_c$, and $F$, respectively. *p*-values correspond to Kruskal–Wallis tests.

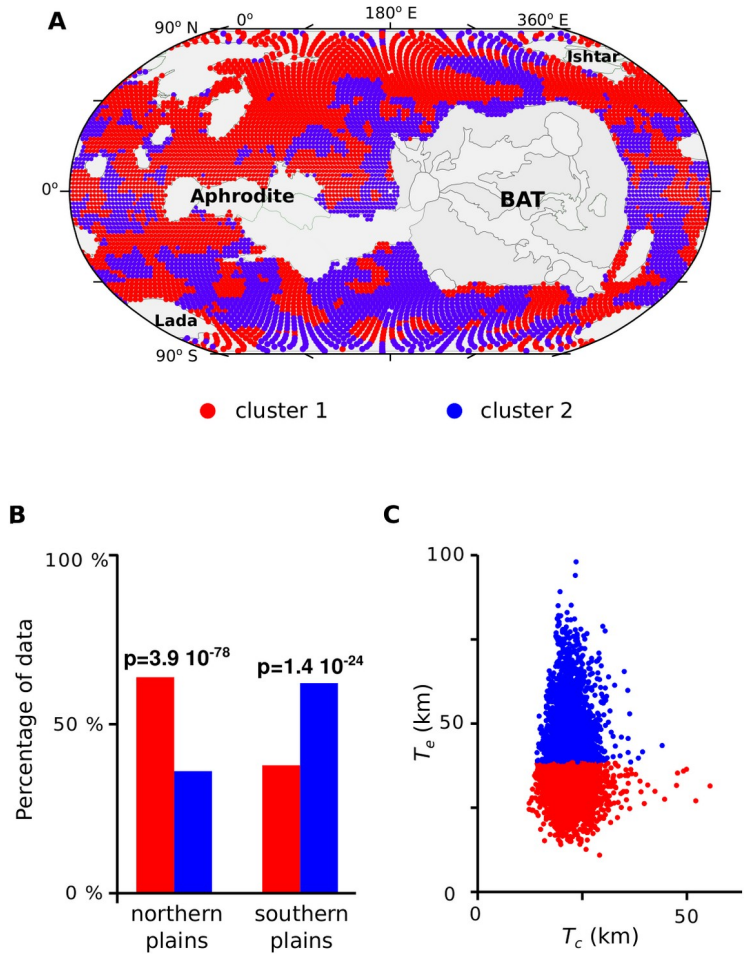

**Fig. 3. Statistical fuzzy model separating data from the volcanic plains out of BAT.** (**A**) Spatial distribution of clusters shown in a global Robinson projection map; data points are homogeneously distributed across the area. (**B**) The bar plots show the percentage of data in each cluster, comparing the northern and southern plains provinces. *p*-values were obtained through chi-square tests. Note the dominance of Cluster 1 in the northern plains and Cluster 2 in the southern plains. (**C**) A scatter plot indicates that the model autonomously selected $T_e$ as the main discriminating parameter between the clusters.

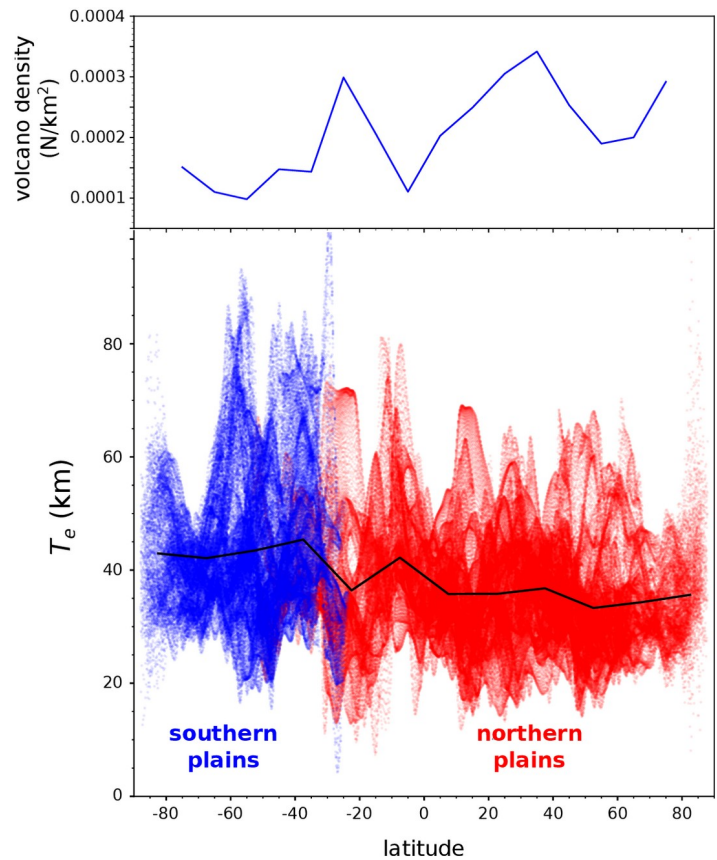

**Fig. 4. Variation of $T_e$ and volcano spatial density with latitude showing the difference between the northern and southern plains.** The variation of the effective elastic lithospheric thickness ($T_e$) of the volcanic plains outside the BAT area with latitude is shown by color points. Red points correspond to northern plains and blue points to southern plains. The average $T_e$ value is indicated by the black line calculated using 15°-latitude bins. The blue line shows the volcano spatial density in number of volcanoes smaller than 5 km per km$^2$ using the volcano database of (38). A strong variation of both parameters around 30º S takes place where we establish the limit between southern and northern plains. Note that a relatively thinner lithosphere is associated with a higher volcano spatial density.

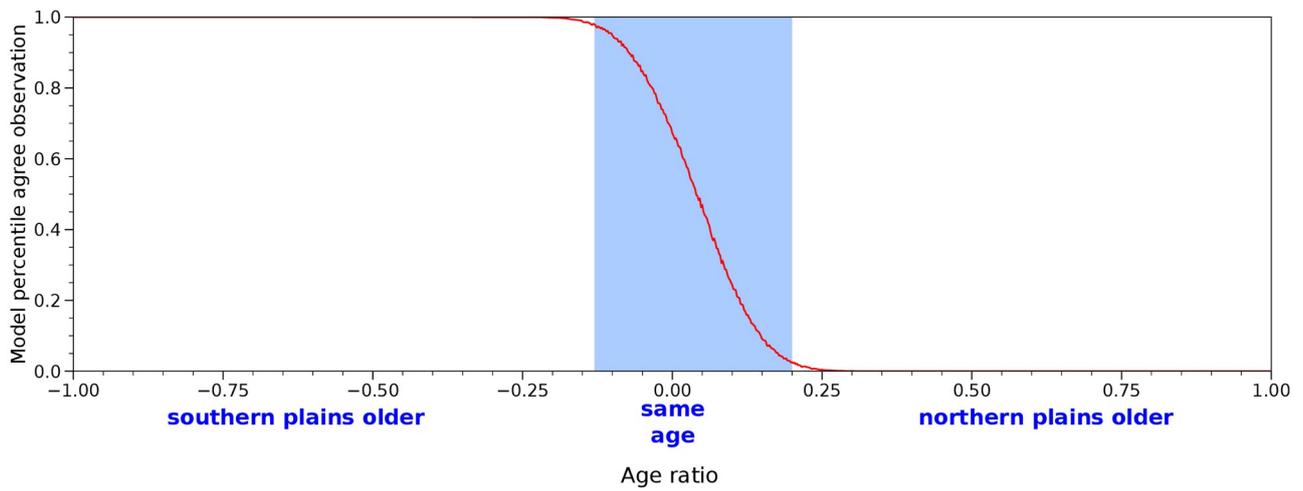

**Fig. 5. Cratering simulations of northern and southern plains.** Results of 10,000 simulations of random cratering in space and time for each of 2,000 cases (horizontal axis) of varying age differences between the northern and southern plains. An age ratio of 0 indicates both units have the same age; an age ratio of 0.15 indicates that the southern plains are 15% younger than those in the north. Under the red curve, the simulations have more craters in the southern plains than are observed and fewer in the north. Above the red curve the opposite is true. The percentile of simulations that agree with actual crater observations is indicated by the red curve (i.e. a value of 0.8 on the vertical axis means that, for a given age difference, 80% of 10,000 simulations show more craters than the observed value in the southern plains and fewer in the northern plains, while the remaining 20% show the opposite). The blue box indicates the age ratio range (-0.13 to 0.20) allowed by crater statistics for a 95% confidence interval.

**CORRESPONDING AUTHOR**

Ignacio Romeo (iromeobr@ucm.es).

**AUTHOR CONTRIBUTIONS**

I.R. designed the research, constructed the global geological map, classified the data by geological provinces, and analyzed the results. A.J.D. and J.F.K. prepared the lithospheric data for the analysis. M.M.E. conducted the statistical tests and developed the statistical model. R.M.H. and P.K.B. provided the volcano database. I.R. performed the cratering tests. I.R., J.R., A.J.D., M.M.E., I.E.G. and J.A.L interpreted the results of the analysis. I.R. wrote the first draft of the manuscrip. M.M.-E. wrote the statistical sections. J.R. and P.K.B. improved the text. Figure design and preparation were carried out by I.R., M.M.-E. and A.J.D. All authors discussed the results and contributed to the final version of the manuscript. I.R. and J.R. secured funding.

**ACKNOWLEDGEMENTS**

This work was supported by the Spanish Agencia Estatal de Investigación through the research project PID2022-140686NB-I00 (MARVEN) and grant PR3/23-30839 (GEOMAVE), funded by the Universidad Complutense de Madrid.

**COMPETING INTERESTS**

The authors declare no competing interests.

**DATA AVAILABILITY**

The database containing crustal thickness, effective elastic thickness and subsurface load fraction and the global map of geological provinces, are publicly available at https://doi.org/10.5281/zenodo.15719877.

.